\documentclass[preprint,12pt,useAMS]{elsarticle}

\usepackage{amssymb}
\usepackage{amsmath}

\journal{Physics of the Dark Universe}

\begin{document}

\begin{frontmatter}



\title{Fitting dark matter mass with the radio continuum spectral data of the Ophiuchus cluster}


\author{Man Ho Chan, Chak Man Lee}

\address{Department of Science and Environmental Studies, The Education University of Hong Kong \\ 
Tai Po, New Territories, Hong Kong, China}

\ead{chanmh@eduhk.hk}

\begin{abstract}
Recent gamma-ray and anti-proton analyses suggest that dark matter with mass $m=48-67$ GeV annihilating via $b\bar{b}$ channel can explain the Galactic center GeV gamma-ray excess and the anti-proton excess as measured by AMS-02 simultaneously. In this article, by differentiating the contributions of dark matter annihilation and normal diffuse cosmic rays, we show that dark matter with mass $m=40-50$ GeV annihilating via $b\bar{b}$ channel with the thermal relic annihilation cross section can best explain the radio continuum spectrum of the central radio halo of the Ophiuchus cluster. This mass range, annihilation cross section, and the annihilation channel is completely consistent with the dark matter interpretations of the GeV gamma-ray excess and the anti-proton excess.
\end{abstract}

\end{frontmatter}



\section{Introduction}
Recently, some analyses of Galactic gamma rays and anti-protons suggest that dark matter annihilation via $b\bar{b}$ channel might be able to explain the Galactic center GeV gamma-ray excess \cite{Daylan,Calore} and the anti-proton excess as measured by AMS-02 \cite{Cholis}. The best-fit ranges of dark matter mass and annihilation cross sections suggested are surprisingly overlapped. Combining the two results gives $m=48-67$ GeV and $\sigma v=(1.4-2.4) \times 10^{-26}$ cm$^3$ s$^{-1}$ \cite{Cholis}. Interestingly, the best-fit annihilation cross section is close to the thermal relic annihilation cross section predicted by standard cosmology $\sigma v=2.2 \times 10^{-26}$ cm$^3$ s$^{-1}$ (for $m \ge 10$ GeV) \cite{Steigman}. Besides, this small range of dark matter mass just satisfies the stringent limits of the Fermi-LAT gamma-ray observations of dwarf galaxies \cite{Albert,Cholis}.
 
To independently verify the above claims, we need another kind of observation. In fact, some studies have explored the possibility of using observational data of galaxy clusters to constrain dark matter properties. For example, the study in \cite{Colafrancesco} analyses the radio data of Coma cluster and obtain some constraints for dark matter. Later, the study in \cite{Storm} analyses the radio data of several galaxy clusters to give some more stringent constraints. Using the observational data of galaxy clusters is good because the contribution of pulsars is not very important for a large region of interest (ROI $>100$ kpc). However, the contribution of normal diffuse cosmic rays in galaxy clusters is quite significant. Most of the previous studies have not taken this contribution into account. If one can differentiate the contributions of dark matter annihilation and the normal cosmic rays, the limits of dark matter would be much more stringent. 

In this article, we analyse the archival data of a large galaxy cluster, the Ophiuchus cluster, and constrain the mass of thermal relic annihilating dark matter. By using the radio continuum spectral data of a central halo, we can differentiate the contributions of cosmic rays and dark matter annihilation so that we can better constrain the properties of dark matter. We show that thermal relic dark matter with $m \approx 40-50$ GeV annihilating via $b\bar{b}$ channel can best fit the radio continuum spectrum of the Ophiuchus cluster. This mass range is completely consistent with the dark matter interpretations of the GeV gamma-ray excess and the anti-proton excess. 

\section{The dark matter annihilation model}
Dark matter annihilation would give a large amount of high-energy electrons and positrons. The energy spectra of electrons or positrons for different annihilation channels are well-known \cite{Cirelli}. These high-energy electrons and positrons would emit synchrotron radiation when there is a strong magnetic field, which could be detected by radio telescopes. For very low redshift clusters, the average synchrotron power at frequency $\nu$ is given by \cite{Storm}
\begin{equation}
P_{\rm syn}=\int_0^\pi d\theta \frac{(\sin \theta)^2}{2} 2\pi \sqrt{3}r_em_ec\nu_gF_{\rm syn} \left(\frac{x}{\sin \theta} \right),
\end{equation}
where $\nu_g=eB/(2\pi m_ec)$, $B$ is the magnetic field strength, $r_e$ is the classical electron radius, and the quantities $x$ and $F_{\rm syn}$ are defined as
\begin{equation}
x= \frac{2 \nu}{3 \nu_g \gamma^2} \left[1+ \left(\frac{\gamma \nu_p}{\nu} \right)^2 \right]^{3/2},
\end{equation}
where $\gamma$ is the Lorentz factor of the electrons or positrons and $\nu_p=8890[n(r)/1~{\rm cm}^{-3}]^{1/2}$ Hz is the plasma frequency ($n(r)$ is the local thermal electron density), and
\begin{equation}
F_{\rm syn}(y)=y \int_y^{\infty} K_{5/3}(s)ds \approx 1.25y^{1/3}e^{-y}(648+y^2)^{1/12}.
\end{equation}

When high-energy electrons and positrons are produced via dark matter annihilation, they would cool down after diffusion. The cooling and diffusion of electrons and positrons can be modeled by the following diffusion equation \cite{Atoyan}:
\begin{equation}
\frac{\partial f}{\partial t}=\frac{D(E)}{r^2} \frac{\partial}{\partial r} \left(r^2 \frac{\partial f}{\partial r} \right)+ \frac{\partial}{\partial E} \left[b(E)f \right]+Q,
\end{equation}
where $f$ is the energy density spectrum of electrons or positrons, $D(E)$ is the diffusion coefficient function, $b(E)$ is the cooling function and $Q$ is the source function. Generally speaking, the diffusion of high-energy electrons and positrons is not important in galaxy clusters \cite{Storm}. The diffusion length scale of an electron with initial energy $E$ is given by \cite{Yuan}
\begin{equation}
\lambda \sim 3.7~{\rm kpc} \left( \frac{D_0}{\rm 10^{28}~cm^2/s} \right)^{1/2} \left( \frac{\omega_0}{\rm 1~eV/cm^3} \right)^{-1/2} \left(\frac{E}{\rm 10~GeV} \right)^{-1/3},
\end{equation}
where $D_0$ is the diffusion coefficient and $\omega_0 \sim 1$ eV/cm$^3$ is the total radiation energy density in a galaxy cluster. Even if we assume a conservative diffusion coefficient $D_0=10^{29}$ cm$^2$ s$^{-1}$, the diffusion length scale is $\lambda \sim 10$ kpc, which is much smaller than the size of a typical radio halo ($\ge 100$ kpc) in a galaxy cluster. Therefore, we can safely neglect the diffusion term in Eq.~(4). In equilibrium state, the solution is
\begin{equation}
f=\frac{dn_e}{dE}=\frac{(\sigma v)[\rho(r)]^2}{2m^2b(E)} \int_E^m \frac{dN_{e,inj}}{dE'}dE',
\end{equation}
where $\rho(r)$ is the dark matter density and $dN_{e,inj}/dE'$ is the injection energy spectrum of dark matter annihilation. The cooling of electrons and positrons is dominated by four processes, synchrotron, inverse Compton scattering, bremsstrahlung and Coulomb losses. The cooling function (in $10^{-16}$ GeV s$^{-1}$) is given by \cite{Colafrancesco}
\begin{equation}
\begin{aligned}
b(E)
=&0.0254E^2B^2+0.25E^2+1.51n(r)\left[0.36+\log \left(\frac{\gamma}{n(r)} \right) \right]
\\
& +6.13n(r) \left[1+\frac{1}{75} \log \left(\frac{\gamma}{n(r)} \right) \right],
\end{aligned}
\end{equation}
where $n(r)$, $E$ and $B$ are in the units of cm$^{-3}$, GeV and $\mu$G respectively. 

The thermal electron number density $n(r)$ is usually modeled by a single-$\beta$ profile \cite{Chen}:
\begin{equation}
n(r)=n_0 \left(1+ \frac{r^2}{r_c^2} \right)^{-3\beta/2},
\end{equation}
where $n_0$, $r_c$ and $\beta$ are empirical parameters obtained from the surface brightness profile (see Table 2). For the magnetic field strength, theoretical models indicate that its profile in a galaxy cluster traces the electron number density \cite{Govoni}. It is commonly modeled by the following form \cite{Storm,Govoni}:
\begin{equation}
B(r)=B_0 \left[\left(1+\frac{r^2}{r_c^2} \right)^{-3\beta/2} \right]^{\eta},
\end{equation}
where $B_0$ is the central magnetic field and $\eta=0.5-1.0$ is the index modeled in simulations. 

For galaxy clusters, the density of dark matter can be described by the Navarro-Frenk-White (NFW) profile \cite{Navarro}. However, there is no robust determination of the scale radius and scale density for each galaxy cluster. Some studies determine these parameters by using the virial mass $M_{200}$ and the virial radius $R_{200}$ of a galaxy cluster \cite{Sanchez}. However, the determinations of the virial mass and virial radius involve the details of the outskirt region of a galaxy cluster, which is usually very uncertain \cite{Reiprich}. Furthermore, we need to assume a power-law form of mass-concentration relation which also contains some systematic uncertainty.  

Fortunately, there is another way to model the dark matter density profile in a galaxy cluster. For a non-cool-core cluster, the temperature of hot gas is nearly constant \cite{Hudson} and the gravitational mass profile $M(r)$ can be probed by using hydrostatic equilibrium:
\begin{equation}
\frac{dP}{dr}=-\frac{GM(r) \rho_g}{r^2}
\end{equation}
where $P=\rho_gkT/(\mu m_p)$ is the pressure of the hot gas, $\rho_g$ is the hot gas density profile, $\mu=0.59$ is the molecular weight and $T$ is the hot gas temperature. Using the single-$\beta$ model in Eq.~(8) for the hot gas density profile, we get 
\begin{equation}
M(r)=\frac{3kT\beta r^3}{\mu m_pG(r^2+r_c^2)}.
\end{equation}
Since the dark matter density dominates the density of a galaxy cluster, the dark matter density is given by
\begin{equation}
\rho(r) \approx \frac{1}{4\pi r^2} \frac{dM(r)}{dr}=\frac{3kT \beta}{4\pi G\mu m_p} \left[ \frac{3r_c^2+r^2}{(r^2+r_c^2)^2} \right].
\end{equation}
Although some studies have pointed out that the mass profile determined may have $\sim 10-20$\% systematic uncertainty (the hydrostatic bias) \cite{Biffi}, we still follow this way to model the dark matter density because it involves the minimum uncertainty.

By combining all the above equations, we can get the radio flux density emitted by a galaxy cluster due to dark matter annihilation:
\begin{equation}
S_{DM}(\nu)=\frac{1}{4\pi D_L^2} \int_0^R \int_{m_e}^m 2 \frac{dn_e}{dE}P_{\rm syn} dE(4\pi r^2)dr.
\end{equation}
The factor 2 in the above equation indicates the contributions of both high-energy positrons and electrons. Here, we assume that the dark matter distribution is spherically symmetric and the distance to the galaxy cluster $D_L$ is large so that it is close to a point-source emission.  

\section{Spectral data fitting}
Generally speaking, the total radio flux density $S(\nu)$ contributions can be divided into two components: thermal contribution $S_{th}(\nu)$ and non-thermal contribution $S_{nth}(\nu)$. The thermal contribution can be calculated by 
\begin{equation}
S_{th}(\nu)=\frac{1}{4\pi D_L^2} \int_0^R 4\pi \kappa(\nu,T)r^2dr,
\end{equation}
where $\kappa(\nu,T)$ is the spectral emissivity, which can be written as \cite{Longair}
\begin{equation}
\begin{aligned}
\kappa(\nu,T)
& =6.8 \times 10^{-42}\left( \frac{T}{10^8~\rm K} \right)^{-1/2}g(\nu,T)[n(r)]^2 
\\
& \times \exp \left(-\frac{h \nu}{kT} \right)~{\rm erg~s^{-1}~cm^{-3}~Hz^{-1}},
\end{aligned}
\end{equation}
where $g(\nu,T) \sim 10$ is the gaunt factor. 

For non-thermal contribution, it consists of dark matter annihilation contribution $S_{DM}$ and normal cosmic-ray contribution $S_{CR}$. The contribution $S_{DM}$ can be calculated by Eq.~(13) while the contribution $S_{CR}$ is difficult to model. The discrepancies between theoretical predictions and observations for cosmic rays are pretty large. Some predicted non-thermal emission resulting from hadronic cosmic ray interactions in the intra-cluster medium exceeds observational radio and gamma-ray data \cite{Jacob}. Nevertheless, numerical simulations show that the spectral index for GeV cosmic rays is close to a constant \cite{Nava}. For the simplest diffusion model of cosmic rays, this would give a constant spectral index $\alpha_{CR}$ for radio continuum spectrum \cite{Jaffe}. However, some more complicated mechanisms might change the spectral index at high frequencies. Based on the studies of the diffuse radio emission in the Coma cluster, the emission mechanisms can be classified into three categories \cite{Thierbach}: 1. primary electron emission \cite{Jaffe,Rephaeli,Rephaeli2}, 2. secondary electron emission \cite{Dennison}, and 3. in-situ model \cite{Jaffe,Roland}. For some primary electron emission and secondary electron emission models, the spectral index is constant for all frequencies and there are only two free parameters involved \cite{Thierbach}. For the other models, the spectral index is nearly a constant in the low frequency regime only. The magnitude of the spectral index would increase in the high frequency regime and the number of free parameters involved is three \cite{Thierbach} (for a review, see \cite{vanWeeren}). 

For the radio halo that we are going to analyse (the radio halo in the Ophiuchus cluster, see below), the spectral index is very close to a constant. Therefore, we simply assume the constant spectral index model to minimize the involved free parameters. In fact, based on the radio spectra in many galaxies, we also note that their spectral index are very close to constant \cite{Srivastava,Mulcahy}. The cosmic-ray contributions are usually modelled by a simple power law $S_{CR}=S_{CR,0}\nu^{-\alpha_{CR}}$. Therefore, the non-thermal radio flux density emitted by a galaxy cluster can be explicitly written as
\begin{equation}
S_{nth}(\nu)=S_{DM}(\nu)+S_{CR,0}\nu^{-\alpha_{CR}},
\end{equation}
where $\nu$ is in GHz. The contribution of cosmic rays is very significant. Previous studies usually ignore this contribution so that the resultant constraints for dark matter are less stringent. Using the radio continuum spectrum can differentiate the contributions of cosmic rays and dark matter annihilation. It is because the spectral index for dark matter annihilation and cosmic rays could be different. For example, if dark matter annihilating via non-leptophilic channels (e.g $b\bar{b}$ channel), its contribution dominates in the lower frequency regime while cosmic-ray contribution dominates in the higher frequency regime. If we can get a radio continuum spectrum with a wide frequency range, it is possible for us to differentiate these different contributions to get more stringent constraints of dark matter.

In the following analysis, we use the archival radio data of the Ophiuchus cluster \cite{Murgia}. The Ophiuchus cluster is a dark-matter-rich large cluster which contains several large radio halos. We choose the central halo to constrain dark matter because it has a higher dark matter density. The radius of the central radio halo is $R=250$ kpc \cite{Murgia,Jacob}. The radio continuum spectral data are shown in \cite{Murgia}. The frequency range ($\nu=0.153-1.477$ GHz) is wide enough to do the analysis and the spectral index is very close to a constant. The thermal contribution $S_{th}(\nu)$ is of the order $10^{-5}$ Jy, which is nearly negligible. Therefore, the non-thermal contribution $S_{nth}$ is equal to the total observed radio flux density (see Table 1). In Table 2, we show all of the related parameters used in the analysis. The uncertainty for each parameter is less than 10\%. 

Besides, the Ophiuchus cluster is classified as a non-cool-core cluster in \cite{Chen}. The emission measure weighted temperature $T_m=10.26$ keV is very close to the halo temperature $T_h=10.25$ keV in the Ophiuchus cluster, where $T_h$ was determined by accepting a small, second lower-temperature component, to allow for a low temperature phase in a possible cooling core in the central cluster region \cite{Chen}. This means that it does not have a cool core temperature which is significantly lower than the mean temperature. However, using other criteria such as the central value of the entropic function, some studies classify the Ophiuchus cluster as a cool-core cluster \cite{Murgia,Jacob}. Following the study in \cite{Hudson}, it shows that using central cooling time is the best parameter for low redshift clusters to segregate cool-core and non-cool-core clusters. For the central cooling time smaller than $7.7$ Gyr, those clusters would be classified as cool-core clusters \cite{Hudson}. The central cooling time of the Ophiuchus cluster is $14.4^{+1.0}_{-0.8}$ Gyr \cite{Chen}, which is significantly larger than 7.7 Gyr. Therefore, based on the criterion set in \cite{Hudson}, it should be regarded as a non-cool-core cluster. Furthermore, in \cite{Kunz}, the Ophiuchus cluster is also regarded as a non-cool-core cluster for the consideration of the magnetic field model which we are going to use. In fact, no matter the Ophiuchus cluster is regarded as a non-cool-core cluster or not, the fact $T_m \approx T_h$ shows that using a constant temperature profile for the Ophiuchus cluster in the hydrostatic equilibrium calculations is a very good approximation to model the dark matter density for the central region.

The only uncertainty is the magnetic field strength. The central magnetic field strength $B_0$ and the index $\eta$ are not very certain. Theoretical models suggest that $\eta$ should be 0.5 and $B_0$ can be calculated by \cite{Kunz,Govoni}
\begin{equation}
B_0=11 \epsilon^{-1/2} \left(\frac{n(r)}{0.1~\rm cm^{-3}} \right)^{1/2} \left(\frac{T}{2~\rm keV} \right)^{3/4}~\mu G,
\end{equation}
where $\epsilon=0.5-1$. This gives $B_0 \approx 9.5-13$ $\mu$G. On the other hand, based on the observational results, the study in \cite{Govoni} suggest that the central magnetic field strength exhibits a power law of the central electron density $B_0 \propto n_0^{0.47}$. If we follow this power law to determine $B_0$ for the Ophiuchus cluster, we get $B_0 \approx 7$ $\mu$G. Therefore, using $B_0=7-13$ $\mu$G and $\eta=0.5-1$ \cite{Govoni} will be appropriate to do the analysis. 

For each dark matter mass, we minimize the $\chi^2$ values by varying $S_{CR,0}$ and $\alpha_{CR}$. It is defined as
\begin{equation}
\chi^2=\sum_i \frac{(S_i-S_{nth,i})^2}{\sigma_i^2},
\end{equation}
where $S_i$ is the observed radio flux and $\sigma_i$ is the uncertainty of observational data. We follow the standard cosmology to use the thermal relic annihilation cross section $\sigma v=2.2 \times 10^{-26}$ cm$^3$ s$^{-1}$ (for $m \ge 10$ GeV) \cite{Steigman}. In Fig.~1, we show the $\chi^2$ values as a function of dark matter mass for two extreme values of $B_0$ and $\eta$ ($b\bar{b}$ channel). We can see that the difference in the positions of the minimum $\chi^2$ values is very small for varying the parameters of $B_0$ and $\eta$. The best-fit mass for $b\bar{b}$ channel is $m=40-50$ GeV. In Fig.~2, we show the $\chi^2$ values of four popular channels ($e^+e^-$, $\mu^+\mu^-$, $\tau^+\tau^-$ and $b\bar{b}$) using $B_0=13$ $\mu$G and $\eta=0.5$. We find that only $e^+e^-$ and $b\bar{b}$ channels have $\chi^2$ troughs and both are located at $m=50$ GeV. However, only the $\chi^2$ trough for $b\bar{b}$ is clearer and deeper, with $\chi^2=5.2$ at $m=50$ GeV (the smallest $\chi^2$ value for $e^+e^-$ is 6.7). The corresponding best-fit radio spectrum is shown in Fig.~3. The best-fit parameters for $e^+e^-$ and $b\bar{b}$ channels are shown in Table 3. Apart from the range of the best-fit dark matter mass, we can also see that the effect of the magnetic field parameters for the $b\bar{b}$ channel on $S_{CR,0}$ and $\alpha_{CR}$ is small (see Table 3). However, for the $e^+e^-$ channel, the possible ranges of $S_{CR,0}$ and $\alpha_{CR}$ are quite large. This is expected because more electron and positron pairs would be produced via leptophilic channels like $e^+e^-$ channel. Therefore, the variation of magnetic field parameters are more sensitive for the $e^+e^-$ channel. 

Generally speaking, the cosmic-ray spectral index $\alpha_{CR}$ depends on the injected spectra of the cosmic-ray sources. These sources include AGNs and supernovae. The possible range of the sources' spectral index is very large, which could range from $\sim 0$ to $\sim 1.5$ \cite{Pei,Randall,Zhang}. The combination of these sources' contributions would give the final spectral index $\alpha_{CR}$. In particular, the sources like AGNs can be roughly divided into three different categories according to their spectral index: 1. ultra-steep spectrum sources (USS) ($\alpha_{CR}=1.0-1.5$), 2. steep spectrum sources (SSS) ($\alpha_{CR}=0.5-1.0$), and 3. flat spectrum sources (FSS) ($\alpha_{CR}=0.0-0.5$) \cite{Zhang}. Statistical studies show that more than half of the sources are SSS \cite{Zhang}. Our best-fit spectral index (the smallest $\chi^2$) for $e^+e^-$ and $b\bar{b}$ are 1.19 and 0.22 respectively. Although these values are not within the most probable range ($0.5-1.0$), it is still quite probable to have these values of the resultant spectral index.

For $m$ larger than 100 GeV, the dark matter contribution becomes very small so that the cosmic-ray contribution dominates the radio spectrum. Therefore, the $\chi^2$ values will converge to constant. Without dark matter contribution, the converged value is $\chi^2 \approx 7.5$, which is somewhat larger than the best-fit value $\chi^2=5.2$. Although the difference is not very large, it shows that including dark matter contribution would give a better explanation to the radio spectral data. Therefore, we conclude that dark matter annihilating via $b\bar{b}$ channel with $m=40-50$ GeV and the thermal relic cross section can best explain the observed radio spectrum of the central radio halo of the Ophiuchus cluster. On the other hand, for $m$ is small (e.g. $m<40$ GeV), lowering the magnetic field strength would give a better fit because the cosmic-ray contribution is more dominant (see Fig.~1).  

In the above analysis, we did not include any substructure contributions. Simulations show that some substructures exist in dark matter profile which can boost the annihilation signals \cite{Gao,Anderhalden,Marchegiani}. However, no observational evidence of these substructures is found. Also, the systematic uncertainty of modelling the substructure contribution is quite large \cite{Bartels} and the effect of the substructure emission near the central region of the Ophiuchus cluster is not very significant. Therefore, we neglect the substructure contributions in our analysis. 

\begin{figure}
\vskip 10mm
 \includegraphics[width=140mm]{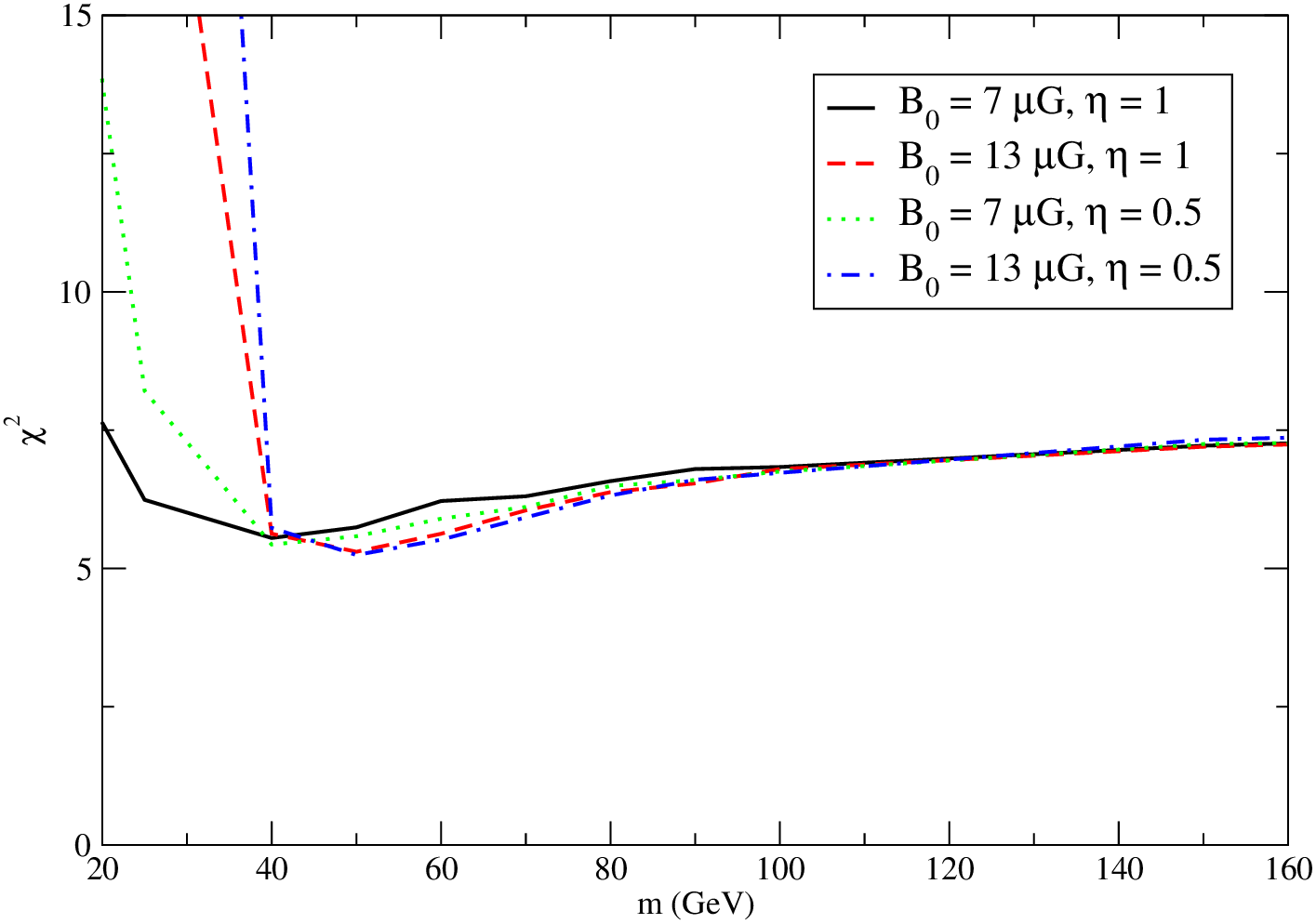}
 \caption{The relation between the $\chi^2$ values and the dark matter mass $m$ for different extreme parameters of $B_0$ and $\eta$ ($b\bar{b}$ channel).}
\vskip 10mm
\end{figure}

\begin{figure}
\vskip 10mm
 \includegraphics[width=140mm]{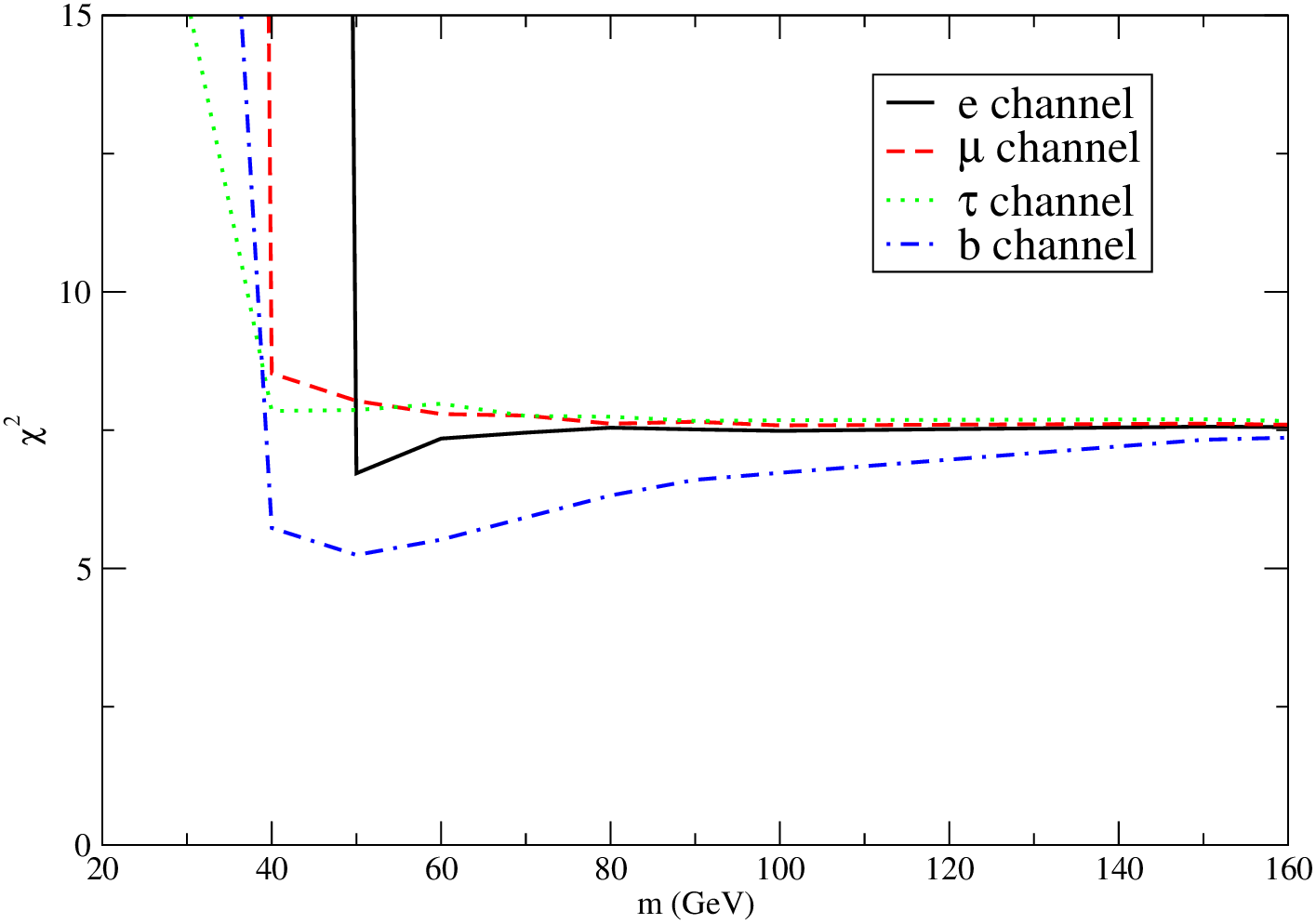}
 \caption{The relation between the $\chi^2$ values and the dark matter mass $m$ for different annihilation channels (assumed $B_0=13$ $\mu$G and $\eta=0.5$).}
\vskip 10mm
\end{figure}

\begin{figure}
\vskip 10mm
 \includegraphics[width=140mm]{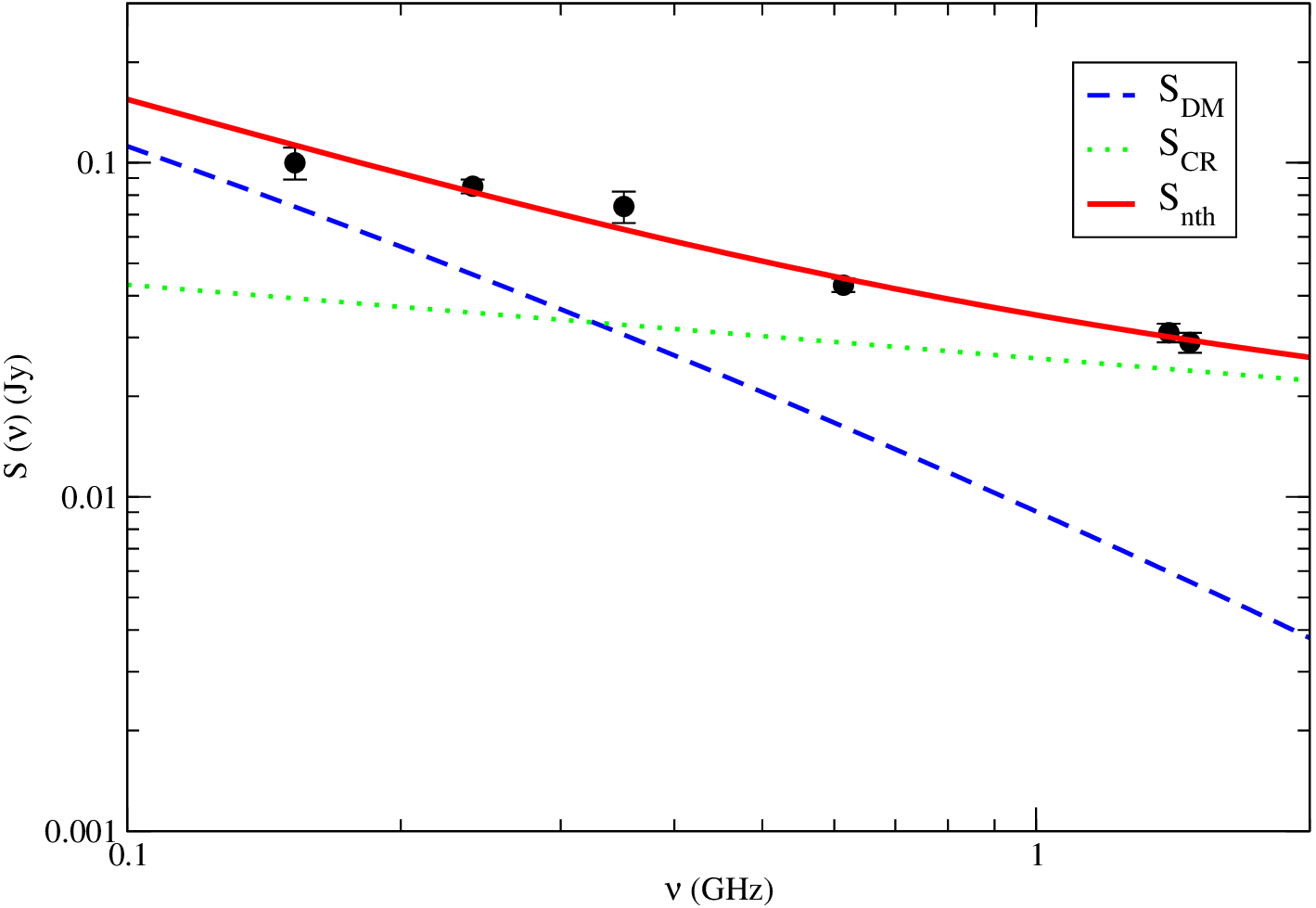}
 \caption{The red solid line is the best-fit radio continuum spectrum ($b\bar{b}$ channel with $m=50$ GeV, $\eta=0.5$ and $B_0=13$ $\mu$G). The corresponding best-fit parameters are shown in Table 3. The blue dashed line and the green dotted line are the dark matter contribution and the cosmic-ray contribution respectively. The observed data are taken from \cite{Murgia}.}
\vskip 10mm
\end{figure}

\begin{table}
\caption{The radio continuum spectral data of the central halo of the Ophiuchus cluster \cite{Murgia}.}
 \label{table1}
 \begin{tabular}{@{}lcc}
  \hline
  $\nu$ (GHz) & $S_{nth}(\nu)$ (Jy) & Uncertainties (Jy) \\
  \hline
  0.153 & 0.100 & 0.011 \\
  0.24 & 0.085 & 0.004 \\
  0.352 & 0.074 & 0.008 \\
  0.614 & 0.043 & 0.002 \\
  1.4 & 0.031 & 0.002 \\
  1.477 & 0.029 & 0.002 \\
  \hline
 \end{tabular}
\end{table}

\begin{table}
\caption{The parameters used in the analysis. The data of $n_0$ and $r_c$ have been re-scaled from \cite{Chen} to match the current Hubble parameter $h=0.68$.}
 \label{table2}
 \begin{tabular}{@{}lll}
  \hline
  Parameter & Observed/fitted value & Reference \\
  \hline
  $n_0$ & $(7.9 \pm 0.5) \times 10^{-3}$ cm$^{-3}$ & \cite{Chen} \\
  $\beta$ & $0.747 \pm 0.035$ & \cite{Chen} \\
  $r_c$ & $204^{+17}_{-15}$ kpc & \cite{Chen} \\
  $T$ & $10.26 \pm 0.32$ keV & \cite{Chen} \\
  $D_L$ & 120 Mpc & \cite{Durret} \\
  \hline
 \end{tabular}
\end{table}

\begin{table}
\caption{The best-fit parameters for $e^+e^-$ and $b\bar{b}$ annihilation channels.}
 \label{table3}
 \begin{tabular}{@{}lcccccc}
  \hline
  Channel & $m$ (GeV) &  $\chi^2$ (d.o.f.=4) & $S_{CR,0}$ (mJy) & $\alpha_{CR}$ & $\eta$ & $B_0$ ($\mu$G) \\
  \hline
  $e^+e^-$ & 50 & 6.7 & 3 & 1.19 & 0.5 & 13 \\
  $e^+e^-$ & 50 & 6.9 & 1 & 1.74 & 1 & 13 \\
  $e^+e^-$ & 60 & 7.4 & 10 & 0.80 & 0.5 & 7 \\
  $e^+e^-$ & 50 & 6.9 & 1 & 1.74 & 1 & 7 \\
  \hline
  $b\bar{b}$ & 50 & 5.2 & 26 & 0.22 & 0.5 & 13 \\
  $b\bar{b}$ & 50 & 5.3 & 27 & 0.26 & 1 & 13 \\
  $b\bar{b}$ & 40 & 5.4 & 28 & 0.26 & 0.5 & 7 \\
  $b\bar{b}$ & 40 & 5.6 & 29 & 0.34 & 1 & 7 \\
  \hline
 \end{tabular}
\end{table}

\section{Discussion}
Previous studies suggest that dark matter annihilating via $b\bar{b}$ channel with $m=48-67$ GeV and $\sigma v=(1.4-2.4) \times 10^{-26}$ cm$^3$ s$^{-1}$ can best explain the Galactic center GeV gamma-ray excess \cite{Daylan,Calore} and the anti-proton excess as measured by AMS-02 \cite{Cholis}. In this article, we show that $40-50$ GeV dark matter annihilating via $b\bar{b}$ channel with the thermal relic cross section $\sigma v=2.2 \times 10^{-26}$ cm$^3$ s$^{-1}$ can best explain the radio continuum spectral data of the central radio halo in the Ophiuchus cluster. Surprisingly, the best range of dark matter mass, annihilation channel and the required annihilation cross section overlap with the previous claims. It means that the dark matter annihilation model with these parameters can simultaneously explain three different high-energy astrophysical phenomena. Note that our results do not rule out the possibility of having $m>50$ GeV. 

In fact, most of the previous studies using galaxy clusters as target objects did not obtain very stringent limits \cite{Storm}. The constrained upper limits of the annihilation cross sections are larger than the thermal relic annihilation cross section. It is because the detected radio signals in galaxy clusters are mainly contributed by diffuse normal cosmic rays from different baryonic processes. In this study, we differentiate the contributions of dark matter annihilation and normal cosmic rays by their respectively different spectral index. This can greatly reduce the upper limit of the dark matter annihilation contribution. Therefore, even if we did not consider the substructure contribution, we can get stringent constraints for thermal relic annihilating dark matter. In other words, using a radio continuum spectrum can help model the contribution of cosmic rays and improve the lower limits of dark matter mass by a significant amount. 

Generally speaking, using the radio continuum spectrum can help detect any alleged signal of dark matter annihilation. Since the spectral index of dark matter annihilation and cosmic rays is somewhat different, the radio continuum spectrum might show a smooth `spectral break' to indicate the possible signal of dark matter annihilation. For the central radio halo of the Ophiuchus cluster, although the $\chi^2$ value with dark matter contribution is smaller than that without dark matter contribution, the radio spectrum does not show a significant `spectral break'. It means that the signal of dark matter annihilation is positive, but not very significant. If the quality of radio data is good enough (i.e. very small observational uncertainties), the `spectral break' might be clear enough to show the signal of dark matter annihilation. Therefore, this may provide another way to search for dark matter indirectly. 

\section{acknowledgements}
The work described in this paper was supported by a grant from the Research Grants Council of the Hong Kong Special Administrative Region, China (Project No. EdUHK 28300518).





\end{document}